\documentclass[reprint,superscriptaddress,showpacs,aps,amssymb,amsmath,amsfonts,pre]{revtex4-1}
\usepackage{hyperref}
\usepackage{epsfig}

\begin{document}
\title{Full statistics of energy conservation in two times measurement protocols}
\author{Tristan Benoist}
\affiliation{Department of Mathematics and Statistics, 
McGill University, 
805 Sherbrooke Street West, 
Montreal,  QC,  H3A 2K6, Canada}
\affiliation{CNRS, Laboratoire de Physique Th\'{e}orique, IRSAMC, Universit\'{e} de Toulouse, UPS, F-31062 Toulouse, France}
\author{Vojkan Jak\v si\' c}
\affiliation{Department of Mathematics and Statistics, 
McGill University, 
805 Sherbrooke Street West, 
Montreal,  QC,  H3A 2K6, Canada}
\author{Annalisa Panati}
\affiliation{Department of Mathematics and Statistics, 
McGill University, 
805 Sherbrooke Street West, 
Montreal,  QC,  H3A 2K6, Canada}
\affiliation{Aix-Marseille Universit\'e, CNRS, CPT, UMR 7332, Case 907, 13288 Marseille, France\\
Universit\'e de Toulon, CNRS, CPT, UMR 7332, 83957 La Garde, France}
\author{Yan Pautrat}
\affiliation{Laboratoire de Math\'ematiques,
Universit\'e Paris-Sud,
91405 Orsay Cedex, France}
\author{Claude-Alain Pillet}
\affiliation{Aix-Marseille Universit\'e, CNRS, CPT, UMR 7332, Case 907, 13288 Marseille, France\\
Universit\'e de Toulon, CNRS, CPT, UMR 7332, 83957 La Garde, France}

\newcommand{\e}{{\rm e}}
\newcommand{\tr}{{\rm tr}}
\renewcommand{\sp}{{\rm sp}}
\newcommand{\cH}{{\mathcal H}}
\newcommand{\cS}{{\mathcal S}}
\newcommand{\cR}{{\mathcal R}}
\newcommand{\cF}{{\mathcal F}}
\newcommand{\ch}{{\mathfrak H}}
\newcommand{\cK}{{\mathcal K}}
\newcommand{\cP}{{\mathcal P}}
\newcommand{\cO}{{\mathcal O}}
\newcommand{\rr}{{\mathbb R}}
\newcommand{\cc}{{\mathbb C}}
\newcommand{\nn}{{\mathbb N}}
\newcommand{\zz}{{\mathbb Z}}
\renewcommand{\i}{{\rm i}}
\newcommand{\Id}{{\rm Id}}
\renewcommand{\d}{{\rm d}}
\renewcommand{\sharp}{\#}

\begin{abstract}
The  first law of thermodynamics states that  the average total energy current between different reservoirs vanishes at large times. In this note we 
examine this fact at the level of the full statistics of two times measurement protocols also known as the Full Counting Statistics. Under  very general conditions, we  establish a tight form 
of the first law asserting that the fluctuations of the total energy current
computed from the energy variation distribution are exponentially suppressed in the large time limit. 
We illustrate this general result using two examples: the Anderson impurity model and a 2D spin lattice model.
\end{abstract}

\maketitle
Recent technical advances in the control of nanoscale systems have enabled the experimental study of out of equilibrium thermodynamics 
in the quantum regime \cite{P15,BLR05,DD11,CINT13,JPAGCJP13,BAPCDL12,TSUMS10,KSSYKSMAP13,KRBMGUIE12}. These new experiments allow for the assessment of 
fluctuations  in addition to the mean heat and particle currents, thus leading to a renewed theoretical investigation of the related quantum thermodynamic laws. 


One of the basic questions in this context concerns the energy flow between two initially isolated large systems $A$ and $B$. The purpose of 
this note is to study some consequences of energy conservation on the
statistical properties of this flow.

By the first law, the average work performed by the
interaction coupling
the two systems is equal to the average
heating of the combined system:
\[
\Delta W_t=\Delta Q_t=\Delta Q_t^A+\Delta Q_t^B.
\label{no-way}
\]
In the case of a sudden switching on/off of the interaction $V$, the
average heating is given by
\begin{equation}
\Delta Q_t=
-\langle V\rangle_t+\langle V\rangle_0
\label{really}
\end{equation}
where $\langle \,\cdot \,\rangle_t$ 
 denotes  the expectation with respect to a suitable system state at time $t$. 
 Whenever $V$ is bounded,  (\ref{really})  gives 
\begin{equation}
\lim_{t\rightarrow \infty}\frac{\Delta Q_t}{t}=0.
\label{sunny}
\end{equation}
The individual
energy currents $\Delta Q_t^{A/B}/{t}$  are also expected  to reach  steady values  $J^{A/B}$. 
They satisfy  $J^A=-J^B$, and are non-vanishing for systems out of equilibrium.  

This note concerns the statistical character of the first law related to the thermodynamics of open quantum systems
at the mesoscopic scale.
Our main result is a refinement of relation 
(\ref{sunny}). It states that the fluctuations 
of the total energy current  are 
exponentially suppressed in the large time limit. 

The nature of  work in quantum physics is more subtle than in classical physics  \cite{talkner}. In the 1990's  Lesovik and Levitov introduced the concept of 
the Full Counting Statistics (FCS) in the study of charge transport \cite{LL}.  The use of the FCS in the definition
of work in quantum physics appeared in the early 2000's in  the works of J. Kurchan and H. Tasaki on the extension of the fluctuation relations to quantum systems \cite{kurchan,Tasaki}. The emerging idea is that in quantum 
mechanics  work should not  be understood as an observable. Instead, the work performed during a given time period is identified with the 
energy variation $\Delta E$ observed in a repeated measurement protocol where the system energy is measured at the beginning and at the end of the period. 
The distribution of the measured energy variation, $\mathbb P_t(\Delta E)$, is the work FCS (we comment on terminology in footnote \footnote{The use of term {\em Counting} in the above context is slightly misleading.  In non-trivial cases, the energy variation (or work) $\Delta E$ is not a discrete quantity in the thermodynamic limit. Nevertheless the name \emph{Full Counting Statistics} is usually used in the literature for the distribution emerging from the repeated measurement protocol we just described.}).
This change of perspective opened a 
whole new area  of research \cite{esposito,campisi}. In particular,  it allowed for the extension of the fluctuation relations to quantum systems \cite{Jarzynski,kurchan, esposito, campisi,JOPP,Tasaki}. 

The  fluctuation relations are intimately related to  the second law of thermodynamics and have been extensively studied \cite{kurchan, Tasaki,esposito, campisi,Crooks,Jarzynski,ES,JOPP}.  Regarding the first law,  the well known identity 
\[{\mathbb E}_t(\Delta E)=\Delta Q_t\]
and \eqref{sunny} give
\begin{equation}
\lim_{t\rightarrow \infty}{\mathbb E}_t\left(\frac{\Delta E}{t}\right)=0
\label{paris-hell}
\end{equation}
where ${\mathbb E}_t$ denotes the expectation with respect 
to  the FCS distribution  ${\mathbb P}_t$ \cite{talkner, JOPP}. 
In this note we sharpen (\ref{paris-hell})  by showing that, under very general conditions, the exponential moment
\[
{\mathbb E}_t\left(\e^{\alpha_m|\Delta E|}\right)
\]
remains bounded as $t\to\infty$
where the constant $\alpha_m>0$ is a measure of the regularity of the interaction $V$ (see (\ref{main}) below).

Until recently, the first law and energy conservation
in the FCS setting have received little
attention in the literature.
In the case where $A$ and $B$ are thermal reservoirs, the FCS of the total energy current was  previously studied 
theoretically in \cite{AGMT}. 
The  works \cite{JPPP,BFJP}
concern the FCS of energy transfer in the thermalization process of a finite level quantum system in contact with a thermal bath, a problem which is 
radically different from the one considered here. We also emphasize that here we are  only interested  in the FCS of the total energy, and 
not in the FCS of the individual energy variations  $\Delta E^{A/B}$. 
We start  with a system described by a finite dimensional Hilbert space ${\cal H}^{(L)}$ where the 
superscript  $L$ refers to the size of the system.  Taking $L\rightarrow \infty$ corresponds to the thermodynamic limit. 
The limiting objects will be denoted without the superscript. 
 Let  $H^{(L)}=H_A^{(L)}+H_B^{(L)}$ be  the Hamiltonian  of the joint but non-interacting system $A+B$. The evolution between the two measurements of $H^{(L)}$ is  generated by $H_V^{(L)}=H^{(L)}+V^{(L)}$, where $V^{(L)}$ denotes
the interaction coupling $A$ and $B$. The initial state is described by the density matrix $\rho^{(L)}$.

 Let 
$P_e^{(L)}$ denote the projection on the eigenspace associated to  the eigenvalue $e$ in the spectrum 
$\sp(H^{(L)})$. The measurement of $H^{(L)}$ at initial time $t=0$ gives $e$ with probability 
$\tr (P_e^{(L)}\rho^{(L)})$. After the measurement  the system  is in the projected state  
\[P_e^{(L)}\rho^{(L)}P_e^{(L)}\big/ \tr (P_e^{(L)}\rho^{(L)}).\]
The second 
measurement of $H^{(L)}$ at a later time $t$ gives $e^\prime$ with probability 
\[
\tr\left(P_{e^\prime}^{(L)}\e^{-i t H_V^{(L)}}P_e^{(L)}\rho^{(L)}P_e^{(L)}\e^{i t H_V^{(L)}}\right)\big/ \tr (P_e^{(L)}\rho^{(L)}).
\]
It follows that the probability of observing the energy variation $\Delta E$ in this  measurement protocol is 
\[
{\mathbb P}_t^{(L)}(\Delta E)=\sum_{e^\prime-e =\Delta E}\tr\left(P_{e^\prime}^{(L)}\e^{-i t H_V^{(L)}}P_e^{(L)}\rho^{(L)}P_e^{(L)}\e^{i t H_V^{(L)}}\right).
\]
The moment generating function of the Full Counting Statistics ${\mathbb P}_t^{(L)}$  is 
\begin{align*}
\chi^{(L)}_t(\alpha)=&
\int_\rr \e^{\alpha\Delta E}\d\mathbb P_t^{(L)}(\Delta E)\\
=&\tr\left(\e^{\alpha H^{(L)}}\e^{-it H_V^{(L)}} \e^{-\alpha H^{(L)}}\tilde \rho^{(L)} \e^{it H_V^{(L)}}\right)
\label{airport}
\end{align*}
where 
\[\tilde \rho^{(L)}=\sum_{e\in \sp(H^{(L)})} P_e^{(L)} \rho^{(L)} P_e^{(L)}.\]

We assume that for 
$\alpha$ purely imaginary, the limit 
\begin{equation}
\lim_{L \rightarrow \infty}\chi_t^{(L)}(\alpha)=\chi_t(\alpha)
\label{airport}
\end{equation}
exists and is a continuous function of $\alpha$. 
 This assumption is harmless and easy to verify in most concrete models of physical interest. 
 By Levy's continuity theorem \cite{gut}, (\ref{airport}) implies that
 the thermodynamic limit $\lim_{L\to \infty} \mathbb P_t^{(L)}=\mathbb P_t$ exists. The probability distribution ${\mathbb P}_t$ is the FCS of the
 thermodynamic system.


Let
\begin{align*}
R^{(L)}(\alpha)=&2|\alpha|\max_{-1\leq s\leq1}\|\e^{s\frac{\alpha}{2} H^{(L)}} V^{(L)}\e^{-s\frac{\alpha}{2} H^{(L)}}\|
\end{align*}
and 
\[
R(\alpha)=\sup_L R^{(L)}(\alpha).
\]
Note that $R(\alpha)$ takes values in $[0, \infty]$ and is an even function. Moreover, $R(\alpha)\geq R(\alpha^\prime)$ 
if $\alpha \geq \alpha^\prime\geq 0$. 
Our regularity condition is that there exists  $\alpha_m>0$ such that 
\begin{equation}
R(\alpha_m)<\infty.
\label{main}
\end{equation}
We emphasize that (\ref{main}) is the only 
regularity assumption we require and that no further hypothesis on the dynamical behaviour of the system is needed. We also 
make no assumptions on the initial state  of the system. 

Our main result is the following strengthening of (\ref{paris-hell}):

{\bf Theorem} For all $t>0$,
\begin{equation}{\mathbb E_t}\left(\e^{\alpha_m |\Delta E|}\right)\leq 2 \e^{R(\alpha_m)}.
\label{montreal-hell}
\end{equation}

An immediate consequence of this result and Chebyshev's inequality \cite{gut} is that for any  $\epsilon >0$, 
\begin{equation}
{\mathbb P}_t\left(\frac{|\Delta E|}{t}\geq \epsilon\right)\leq 2\e^{-t \epsilon \alpha_m + R(\alpha_m)}.
\label{proof}
\end{equation}
Note that if $R(\alpha)<\infty$ for all $\alpha$,  then 
\begin{equation}
\mathbb P_t\left(\frac{|\Delta E|}{t}\geq \epsilon\right)\leq 2 \e^{R(C/\epsilon)-Ct}
\label{strong}
\end{equation}
for any $C>0$. 

The estimates (\ref{proof}) and (\ref{strong}) can be interpreted in terms of the  large deviation theory \cite{DemboZeitouni} (see \cite{BJPPP}). 
For example, (\ref{strong}) implies  that the large deviation rate function of the random variable 
$|\Delta E|/t$  satisfies $I(s)=\infty$ for $s\not=0$, and that  the large deviations are completely suppressed in the large time limit.


The main  novelty of our proof is the derivation of a time independent bound for $\chi_t^{(L)}$ inspired by the bounds proposed in \cite{AGMT}. The derivation is  based on two well-known  inequalities. The first  is  
\[\tr(XY)\leq \|X\|\tr(Y)\]
which holds 
for any two   non-negative matrices $X, Y$. 
The second  states that for any two self-adjoint matrices $T, S$, 
\begin{equation}
\|\e^{T+S}\e^{-T}\|\leq\e^{\max_{0\leq s\leq 1}\|\e^{sT}S\e^{-sT}\|}.
\label{afternoon-hell}
\end{equation}
To prove this inequality, let $\Gamma(s)=\e^{s(T+S)}\e^{-sT}$. Then one has  
\[\partial_s\Gamma(s)=\Gamma(s) \e^{s T} S \e^{-sT},\,\,  \Gamma(0)={\mathbb I}.\]
Using 
\[\|\partial_s \Gamma(s)\|\leq\|\Gamma(s)\|\|\e^{sT}S\e^{-sT}\|\]
 and Gronwall's inequality we obtain
(\ref{afternoon-hell}). 
The bound  (\ref{afternoon-hell}) is similar but unrelated to the bound (3.10) of \cite{lenci05}.

The proof of (\ref{montreal-hell}) proceeds as follows. 
For $\alpha$ real we  set
\[X=\e^{-\frac{\alpha}{2}H_V^{(L)}}\e^{\alpha H^{(L)}}\e^{-\frac{\alpha}{2}H_V^{(L)}}\]
 and 
 \[Y=\e^{-\i t H_V^{(L)}} \e^{\frac{\alpha}{2}H_V^{(L)}}\e^{-\frac{\alpha}{2} H^{(L)}}\tilde\rho^{(L)} \e^{-\frac{\alpha}{2} H^{(L)}}\e^{\frac{\alpha}{2}H_V^{(L)}}\e^{\i t H_V^{(L)}}\] 
(note that $\tilde \rho^{(L)}$ and $H^{(L)}$ commute). Observe that 
\[
\chi_t^{(L)}(\alpha)=\tr (XY)
\]
and that $X, Y$ are non-negative matrices.  We then use the first inequality to derive the estimate 
\[\chi_t^{(L)}(\alpha)\leq \|X\|\tr (Y)\] where 
 \[ \|X\|=\|\e^{-\frac{\alpha}{2}H_V^{(L)}}\e^{\frac{\alpha}{2} H^{(L)}}\|^2
 \]
 and 
 \[\tr (Y)=
\tr \left( \e^{\frac{\alpha}{2}H_V^{(L)}}\e^{-\frac{\alpha}{2} H^{(L)}}\tilde\rho^{(L)} \e^{-\frac{\alpha}{2} H^{(L)}}\e^{\frac{\alpha}{2}H_V^{(L)}}\right).
 \]
The cyclicity of the trace gives 
 \[\tr (Y)=\tr \left(\e^{-\frac{\alpha}{2} H^{(L)}}\e^{\alpha H_V^{(L)}}\e^{-\frac{\alpha}{2} H^{(L)}}\tilde\rho^{(L)}\right).
\]
Applying  the first inequality once again and using that $\tr (\tilde \rho^{(L)})=1$, we derive 
\[
\tr (Y)\leq \|\e^{\frac{\alpha}{2}H_V^{(L)}}\e^{-\frac{\alpha}{2} H^{(L)}}\|^2.
\]
Hence
\[
\chi_t^{(L)}(\alpha)\leq \|\e^{-\frac{\alpha}{2}H_V^{(L)}}\e^{\frac{\alpha}{2} H^{(L)}}\|^2\|\e^{\frac{\alpha}{2}H_V^{(L)}}\e^{-\frac{\alpha}{2} H^{(L)}}\|^2.
\]
Using the second inequality with 
\[T=\mp \frac{\alpha}{2} H^{(L)}, \qquad S=\mp \frac{\alpha}{2} V^{(L)},\]
we obtain
\[
\chi_t^{(L)}(\alpha)\leq \e^{R^{(L)}(\alpha)}.
\]
The regularity assumption (\ref{main}), the existence of the limit (\ref{airport}) for purely imaginary $\alpha$'s, 
 and Vitali's convergence theorem (see Appendix B in \cite{JOPP}) give that for all  complex $\alpha$ with real part ${\rm Re}(\alpha)$ in 
$(-\alpha_m, \alpha_m)$, the limit  $\lim_{L\to\infty} \chi_t^{(L)}(\alpha)=\chi_t(\alpha)$ exists. Moreover, for such $\alpha$'s,  
\[\chi_t(\alpha)=\int_\rr \e^{\alpha\Delta E}\d\mathbb P_t(\Delta E)\]
and 
\[
|\chi_t(\alpha)|\leq \e^{R({\rm Re}(\alpha))}.
\]
It follows that 
\[|\chi_t(\pm \alpha_m)|\leq e^{R(\alpha_m)}.
\]
The last estimate gives 
\[
{\mathbb E}_t\left(\e^{\alpha_m |\Delta E|}\right)\leq\chi_t(-\alpha_m)+\chi_t(\alpha_m)\leq  2 \e^{R(\alpha_m)}
\]
and the theorem follows.

\paragraph{Spin--fermion models.}
Electronic transport through a 1D-lattice containing a single magnetic
impurity is a typical problem involving bounded interactions. The Anderson model \cite{Anderson,Hewson} commonly used to study this question is a specific example of a general class of  spin--fermion models
to which our main theorem applies. 

The study of the FCS of charge transport through the impurities in such models is an active field of research\cite{GK06,GK06b,SGK07,SKG07,SOKT11}. 
We emphasize, however, that here we are only concerned with the statistics of the total energy. 
 
The impurity is described by a quantum dot supporting four different eigenstates:  
empty, occupied by a single electron with either spin up or spin down, or occupied by two electrons with opposite spins. The remaining 
parts of the lattice, regarded as fermionic (say left and right) reservoirs
at different chemical potentials, are described in the tight binding approximation.

Here, the subsystem $A$ is the left side of the lattice together with  the impurity. The lattice right side is the subsystem $B$.

The operator $c_{l/r,\sigma}^\ast(x)$ ($c_{l/r,\sigma}(x)$)
creates (annihilates) an electron with spin $\sigma$ at the lattice site $x$
of the left ($x<0$)/right ($x>0$) reservoir. Similarly, the operator
$d_\sigma^\ast$ ($d_\sigma$) creates (annihilates) and electron with spin $\sigma$ in the dot. 
The anti-commutation
relations $\{c_{l/r,\sigma}(x),c^\ast_{l/r,\sigma'}(x')\}=\delta_{x, x'}\delta_{\sigma,\sigma'}$ and $\{d_\sigma,d_{\sigma'}^\ast\}=\delta_{\sigma,\sigma'}$ hold while the $c$ operators
commute with the $d$ operators.
We use the shorthand $c_{l/r,\sigma}(\phi)=\sum_{x}\overline{\phi}(x)c_{l/r,\sigma}(x)$. The reservoir Hamiltonians are \[H_{l}=\sum_{\sigma=\pm;x,x'<0\atop |x-x'|=1}c^\ast_{l,\sigma}(x)c_{l,\sigma}(x^\prime),\] with  a similar expression for $H_r$.
Let  $h_{l/r}$ be the discrete Laplacian 
of the left/right part of the lattice. Since $h_{l/r}$ is a bounded operator, 
 \[\e^{\alpha H_{l/r}}c_{l/r,\sigma}(\phi)\e^{-\alpha H_{l/r}}=c_{l/r,\sigma}(\e^{\alpha h_{l/r}}\phi)\]
for all real $\alpha$. In particular, for  all $\alpha$, 
\begin{equation}
\|\e^{\alpha H_{l/r}}c_{l/r,\sigma}(\phi)\e^{-\alpha H_{l/r}}\|<\infty.
\label{what}
\end{equation}
The total Hamiltonian is
\[H=H_\cS+H_l+H_r\]
where $H_\cS=\epsilon\sum_{\sigma}d_\sigma^\ast d_\sigma+Ud_+^\ast d_+d_-^\ast d_-$ is the Hamiltonian of the dot. Regarding the subdivision in A/B subsystem, we have $H_A=H_l+H_\cS$ and $H_B=H_r$.
The coupling of the conduction electrons with the dot is described by
$$
V=\sum_{\sigma}\left(d_\sigma^\ast(c_{l,\sigma}(v_{l,\sigma})+c_{r,\sigma}(v_{r,\sigma})) + {\rm h.c.}\right)
$$
for some coupling functions $v_{l/r,\sigma}(x)$.
In the  context of the Anderson model, the superscript $L$ refers to the confinement of  the reservoirs to the finite part of the lattice defined by $|x|\le L$. Such confinement
is necessary to allow for a  meaningful definition of the repeated measurement protocol leading to the FCS. The limit $L\rightarrow \infty$ restores the
extended reservoirs.  
It follows from  relation (\ref{what}) that  $R(\alpha)$ is finite for all $\alpha$, and that our theorem holds for all $\alpha_m>0$.  Hence we have inequality \eqref{strong}:
\[\mathbb P_t\left(\frac{|\Delta E|}{t}\geq \epsilon\right)\leq 2\e^{R(C/\epsilon) - Ct}\]
for any $\epsilon>0$ and any $C>0$.

We also note that one can consider  instead the FCS of $H'=H_l+H_r$ by setting $V'=H_\cS+V$. Then $H'_A=H_l$ and $H'_B=H_r$. One then  obtains 
the same result by replacing $\Delta E$ with  $\Delta E'$. The energy of the impurity is irrelevant in the large time limit. 

\paragraph{Spin systems.}
Another popular class of models involving bounded interactions are  locally interacting spin systems. In \cite{BJPPP} we prove that, under general conditions, our theorem applies to locally interacting spin systems in arbitrary dimension. 
Moreover, for  1D spin systems   with finite range interactions, Araki's results \cite{Araki00} give that $R(\alpha)<\infty$ for 
all $\alpha$, and hence that our  theorem holds for all $\alpha_m>0$. We restrict ourselves to the description of a simple example. 

\begin{figure}
\centering
\includegraphics[scale=0.6]{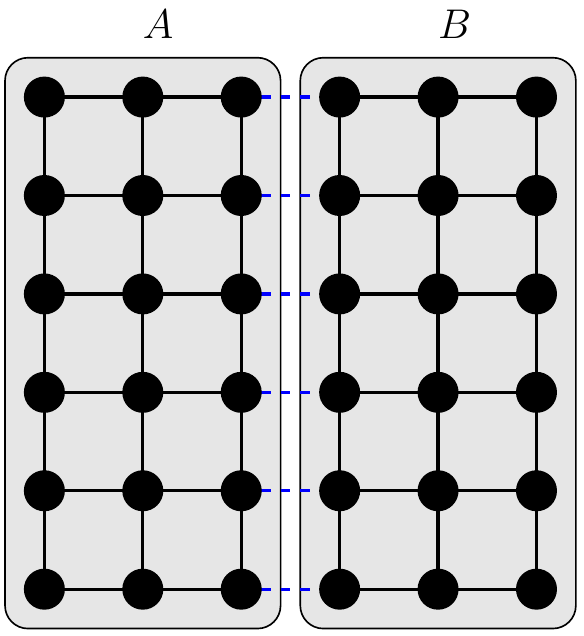}
\caption{A partitioned finite spin system $A+B$. Solid lines represent the
nearest neighbour coupling $J$ and dashed lines the interaction $K_{x,y}$
between the $2$ subsystems.}
\label{Fig1}
\end{figure}

Consider a 2D square lattice of $\frac12$-spins. Let 
$\Lambda_L\subset\zz^2$ be the finite sub-lattice of size $2L\times 2L$.  We denote by $\Lambda_L^\pm$  its left/right half. Subsystems $A$ and $B$ are the spins in $\Lambda_L^-$
and $\Lambda_L^+$ respectively (see Figure~\ref{Fig1}).

The system Hilbert space is $\cH^{(L)}=\bigotimes_{x\in\Lambda_L} \cc^2$.
The Hamiltonian is that of an XY-spin model  where the spins on $\Lambda_L^-$ do not interact with that on $\Lambda_L^+$ \footnote{The matrices $\sigma_x^{(j)}$ act non trivially only on the site $x$ copy of $\cc^2$ with the corresponding Pauli matrix: $\sigma_x^{(j)}=\bigotimes_{y\in\zz^2, x\neq y} \Id \otimes \sigma^{(j)}$.\\$\sigma^{(1)}=\left(\begin{array}{cc}0&1\\1&0\end{array}\right), \sigma^{(2)}=\left(\begin{array}{cc}0&-\i\\\i=&0\end{array}\right),\sigma^{(3)}=\left(\begin{array}{cc}1&0\\0&-1\end{array}\right)$. }:
$$
H^{(L)}=H^{(L,-)}+H^{(L,+)},
$$
with
$$
H^{(L,\pm)}=-\frac{J}{2}\sum_{x,y\text{ nearest}\atop\text{ neighbors in }\Lambda_L^\pm} \left(\sigma_{x}^{(1)}\sigma_y^{(1)} + \sigma_{x}^{(2)}\sigma_y^{(2)}\right),
$$
where $J$ is  a  coupling constant. The interaction is 
\[
V^{(L)}=-\frac{1}{2}\sum_{x\in\Lambda_L^-,y\in\Lambda_L^+} K_{x,y} (\sigma_x^{(1)}\sigma_{y}^{(1)} +  \sigma_x^{(2)}\sigma_{y}^{(2)}),
\]
where
$$
K_{x,y}=\frac\epsilon{1+x_2^2}
$$
if $x=(0,x_2)\in\Lambda_L^-$ and $y=(1,x_2)\in\Lambda_L^+$ 
and $K_{x,y}=0$ otherwise. The boundary between the two halves of the lattice is between the lines $x_1=0$ and $x_1=1$. Note that the 
 interaction intensity decreases as one moves away from  $(0,0)$. An  assumption of this type  is necessary if  $V^{(L)}$ is 
 to remain bounded in the thermodynamic limit $L\rightarrow \infty$. 

For this model one can show that there exists $\alpha_m>0$ such that (\ref{main}) holds  and that our  theorem applies. Hence we have inequality \eqref{proof}:
\[\mathbb P_t\left(\frac{|\Delta E|}{t}\geq \epsilon \right)\leq 2 \e^{-t\epsilon\alpha_m+R(\alpha_m)}\]
for any $\epsilon>0$.

\paragraph{Discussion.}
Under a general condition on the regularity of the  interaction evolution in imaginary time, we have proven a sharp 
form of the first law of thermodynamics for the FCS of energy variation.

Our  result holds for any initial state  of the system.
If one assumes that   systems $A$ and $B$ are initially in  thermal equilibrium at  temperatures $T_A$ and $T_B$, then 
the suppression of the fluctuations of the total energy current can be also proven  by following the arguments of   \cite{AGMT}. 


Under additional assumptions it is possible to deal with cases where
several reservoirs drive the joint system towards a non-equilibrium steady state and to derive properties of the joint distribution of the energy variations in each part of the system.
 A more strict condition on $R(\alpha)$ allows for the generalization of a symmetry of the limiting cumulant generating function proposed in \cite{AGMT}. Combined with time reversal invariance this leads to Onsager's reciprocity relations. We investigate these topics in \cite{BJPPP}.

In the present note we have limited ourselves to bounded interactions. The case of unbounded interactions (an example is the 
spin-boson model) is more technical 
and requires a separate analysis based on an application of Ruelle's quantum transfer operators \cite{JOPP}. Although the physical 
picture emerging from this analysis is of an independent interest, the final results are much less general than in the case of bounded 
interactions  \cite{deroeck, JPPP2}. 

\begin{acknowledgments}
The research of T.B. was partly supported by ANR project RMTQIT (grant ANR-12-IS01-0001-01). T.B.  also wishes to thank Technische Universit\"at M\"unchen and Pr. M. M. Wolf for his hospitality during the last weeks of work on this note. The research of T.B. and Y.P. was partly supported by ANR contract ANR-14-CE25-0003-0. Y.P. also wishes to thank UMI-CRM for financial support, and McGill University for its hospitality. The research of T.B. and V.J. was partly supported by NSERC. The research of A.P. was partly supported by NSERC and ANR (grant 12-JS01-0008-01). The research of C.-A.P has been carried out in
the framework of the Labex Archim\`ede (ANR-11-LABX-0033) and of the A*MIDEX project (ANR-11-IDEX-0001-02), funded by the ``Investisements d'Avenir" French Government programme managed by the French
National Research Agency (ANR).  
\end{acknowledgments}
\bibliography{article_PRB.v1,article}
\end{document}